# Atomic coexistence of superconductivity and incommensurate magnetic order in the pnictide Ba(Fe$_{1-x}$Co$_x$)$_2$As$_2$

Y. Laplace, [1] J. Bobroff, [1*] F. Rullier-Albenque, [2] D. Colson, [2] A. Forget [2]

[1]*Laboratoire de Physique des Solides, Univ. Paris-Sud, UMR 8502 CNRS, 91405 Orsay Cedex, France*
[2]*Service de Physique de l'Etat Condensé, Orme des Merisiers, CEA Saclay (CNRS URA 2464), 91191 Gif sur Yvette cedex, France*

*10 June 2009*

$^{75}$As NMR and susceptiblity were measured in a Ba(Fe$_{1-x}$Co$_x$)$_2$As$_2$ single crystal for x=6%. Nuclear Magnetic Resonance (NMR) spectra and relaxation rates allow to show that all Fe sites experience an incommensurate magnetic ordering below *T=31K*. Comparison with undoped compound allows to estimate a typical moment of *0.05 $\mu_B$*. Anisotropy of the NMR widths can be interpreted using a model of incommensurability with a wavevector $(1/2-\varepsilon, 0, l)$ with ε of the order of 0.04. Below *T$_C$=21.8K*, a full volume superconductivity develops as shown by susceptibility and relaxation rate, and magnetic order remains unaffected, demonstrating coexistence of both states on each Fe site.

PACS numbers : 74.70.-b 74.62.Dh 75.25.+z 76.60.-k

Unconventional superconductivity (SC) has now been evidenced in a wide variety of materials, high Tc cuprates, cobaltates, heavy fermions, organic conductors. Despite their differences, all these compounds share a common unexpected feature: in all their phase diagrams, superconductivity is always adjacent to a long range magnetic ordered phase, usually antiferromagnetic (AF). Deciding whether superconductivity and AF order exclude each other or may coexist has been the subject of intense debates, revived recently by the discovery of Fe-pnictides [1]. These display a spin density wave (SDW) magnetic ordering which turns into a high temperature superconductor when doping is applied [2]. If coexistence were to occur, this could modify the superconducting state itself and put strong constrains on possible theories for superconductivity [3, 4, 5].

This is not the case in the LaOFeAs family, where a first order transition is observed from AF to SC [6]. On the contrary, various macroscopic measurements suggest possible coexistence in the BaFe$_2$As$_2$ family [7], as shown in the phase diagram of fig.1 [8]. However, in the case of K-doping at Ba site, local probes such as muon spin resonance (μSR) [9] or NMR [10] showed that the system segregates into small AF and SC domains, and the resulting phase mimics both types of orders. This type of phase segregation detected by local probes only is indeed a common feature to many of these unconventional superconductors [11]. For Co-doping, situation is not yet settled on atomic scale. A recent μSR study demonstrated full magnetic ordering in a superconducting sample, but superconductivity was evidenced only through resistivity, and the superconducting volume fraction was not reported [12]. In addition, muons are sensitive to the dipolar field of frozen moments up to a few lattice spacings away so nanoscale segregation cannot be ruled out as argued in [13]. In contrast, the $^{75}$As nuclei probed in As NMR are sensitive only to their four near neighbour Fe electrons. This ensures that if $^{75}$As NMR displays both SC and AF, then each corresponding Fe atom is subject to both on *atomic* scale. To our knowledge, the only NMR data suggesting atomic coexistence was reported in CaFe$_2$As$_2$ under pressure [14], but determination of the SC phase remained an open question and SC volume fraction was not determined. We report here a NMR study of high quality Ba(Fe$_{1-x}$Co$_x$)$_2$As$_2$ crystals where, for *x=6 %*, coexistence of SC and SDW is demonstrated at an unpreceedent atomic scale.

Single crystals synthesis and characterization details are reported in [8] together with transport measurements which allowed to establish the phase diagram of fig.1. In order to get good radiofrequency penetration for NMR, we performed our measurements on a crystal divided into a few small pieces, magnetically aligned in epoxy. Echo pulse NMR was measured at *H=7.5T* by sweeping the frequency or at fixed frequency by sweeping the field (the former was prefered in the SC state because field sweep would affect the vortex lattice as well). The quality and homogeneity of the crystals has been evidenced in [8] and is confirmed by a sharp macroscopic superconducting transition (fig.1) and the NMR lineshape and dynamics [20]. The magnetic susceptibility was measured in a SQUID magnetometer in Zero Field Cooled with *H//ab* of *1 G*. We find a critical temperature *T$_C$=21.8 K* and a transition width of about *2.5K*, which implies a maximum Co doping distribution of only *±0.25%* in the whole sample [20]. Demagnetization factors are estimated to be *N$_c$=0.7(1)* and *N$_{ab}$=0.10(5)* for the rectangular shape of our crystal. The volumic susceptibility *χ=-1.15emu/cm$^3$* measured at low *T* corresponds then to a superconducting fraction higher than *95%* and we can safely conclude that our sample exhibits a full volume bulk superconductivity.

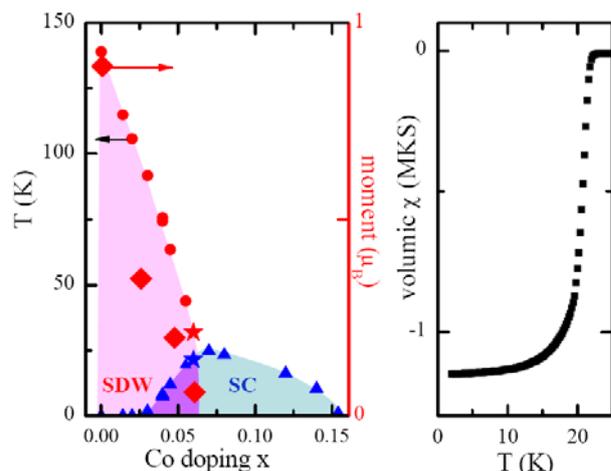

Fig. 1: (Color online) *Left panel:* data for T$_{SDW}$ (circles) and T$_C$ (triangles) for Ba(Fe$_{1-x}$Co$_x$)$_2$As$_2$ taken from Ref.[8], where the stars represent the x=6% composition used here. The moment amplitude

(diamonds) corresponds to the right axis, using values of [15] and this study. *Right panel* : macroscopic susceptibility at H=1G, after Zero Field Cooled.

We now focus on the magnetic ordering as probed by $^{75}$As NMR. The NMR frequency is given by

$$\nu = \frac{\gamma}{2\pi}(H_0 + h_{loc}) + f_q$$

where $\gamma/2\pi=7.2919$ MHz/Tesla is the gyromagnetic ratio of $^{75}$As, $H_0$ is the applied field, $h_{loc}$ is the additional local field experienced by As nuclei due to hyperfine couplings to the Fe electrons, and $f_q$ is due to quadrupolar electric effects. Since the $^{75}$As carries a nuclear spin I=3/2, $f_q$ has three possible values, explaining the three lines observed in $^{75}$As NMR spectra displayed in fig.2.a,b. The very narrow central line at *T=50K* confirms the good homogeneity of our sample and the absence of any Co segregation. In the high temperature paramagnetic regime, $h_{loc}$ is proportional to the spin shift and the electron spin susceptibility, and the values recorded are found consistent with those reported [16]. At low temperature, all three lines strongly broaden, the broadening being much more pronounced for *H//c* than for *H//ab*. This broadening comes from the appearance of large and distributed internal fields $h_{loc}$. This distribution was also observed at smaller Co dopings *x=2%* and *4%* [16]. It signals the appearance of a distribution of the Fe moments amplitude. ). The lineshape alone does not allow to decide the corresponding magnetic pattern, but as shown hereafter, it is the strong anisotropy of the linewidth which allows us to conclude for an incommensurate Spin Density Wave. If it were due instead to a non-periodic disorder induced by Co dopants, such as a variation of the moment amplitude on Co sites, it would not lead to the linewidth anisotropy and would not be observed for *x=2%* as well [16]. This continuous distribution contrasts with the situation encountered in undoped BaFe$_2$As$_2$, where an AF commensurate order leads to only two values of $h_{loc}$ and thus two sets of well splitted lines below T$_{SDW}$ for *H//c*. The T-dependence of the SDW magnetization is monitored by that of the local field distribution proportional to the NMR width. It is displayed in upper panel of fig.3. The onset at *T$_{SDW}$=31K* fits well with the maximum detected in dynamics in fig.3 and is identical to that found using resistivity and Hall effect [8].

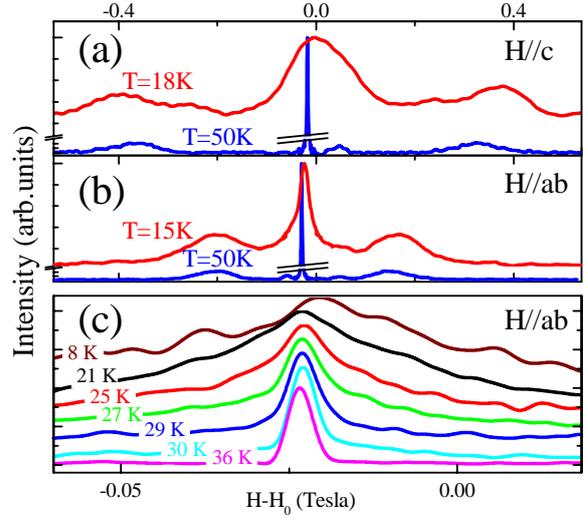

Fig. 2: (Color online) *Panel a and b*: typical NMR spectra for the two orientations of the applied field. A strong broadening is observed below T$_{SDW}$ (a break on the intensity axis was introduced to artificially enlarge quadrupolar satellites at high temperature). *Panel c*: zoom on the H//ab central line for various temperatures (spectra were measured by sweeping the frequency at *H=7.5 Tesla*, and reconverted in field units).

The whole NMR spectrum broadens homogeneously, and we checked that no intensity is lost down to low temperatures, implying that frozen moments develop on all the Fe atoms. If even just a few percent of the sample was not magnetically frozen, one should detect instead a narrow central line on top of the broad distribution as in K-doped BaFe$_2$As$_2$ [10]. This is clearly not the case as shown in both directions of field in fig.2. As already stressed, the very short range of the hyperfine interactions make $^{75}$As NMR sensitive only to its near neighbor Fe, while muons in μSR probe frozen moments far up to a few nm. So *the present NMR experiment allows us to rule out the existence of nanosize segregation, and demonstrates the magnetic ordering at all Fe sites.* Together with the fact that the superconducting fraction is about 100%, our data evidence atomic coexistence of magnetism and superconductivity in the whole sample below T$_C$. Superconductivity is not observed to affect the NMR spectral shape. This is not surprising since the SDW field distribution dominates any static effect due to SC such as vortex field distribution or decrease of the Knight Shift, both only of the order of *10 to 30 G* here.

The typical field distribution *ΔH=2000 G* for *H//c* measured at low temperature (fig.2) is much smaller than the splitting of *30000 G* found in the undoped composition [17]. Such reduction is due to the reduction of the moment itself, and can be used to get an estimate of the moment per Fe. Using *m=0.9 μ$_B$* for *x=0%*, we deduce $m=0.06\ \mu_B$ at $x=6\%$. This value is plotted on the phase diagram of fig.1 together with other values reported in neutron measurements [15].

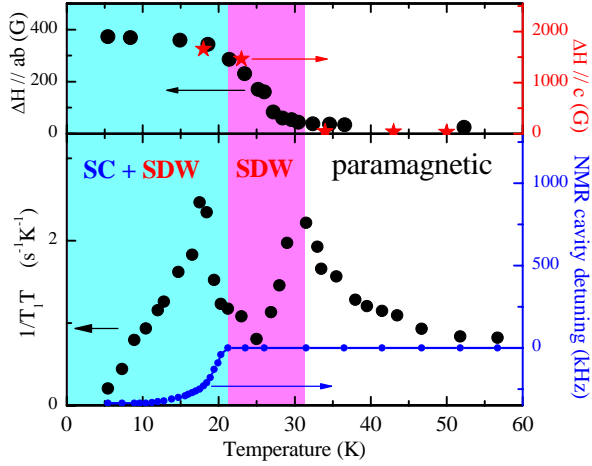

Fig.3 : (Color online) Upper panel : NMR field distribution for the two field orientations shows apparition of the SDW magnetic ordering below $T=31K$. Lower panel : $1/T_1T$ measured for $H//ab$ (left axis) shows two peaks at both SC and SDW transitions, while detuning of the NMR cavity (right axis) demonstrates the apparition of SC.

The incommensurate wavevector can be roughly estimated by comparing the NMR widths for the two directions of applied field. In the undoped case for a commensurate wave vector $(1/2,0,l)$ in reciprocal lattice units, Fe moments are anti-aligned along $a$ axis as displayed in fig.4. This results in a transferred hyperfine field on $^{75}As$ only for $H_0//c$ for symmetry reasons [17]. The observation of a sizeable $\Delta H_{ab}$ for $H//ab$ in our case implies that the Fe moments do not follow this AF pattern anymore. But the small ratio $R = \Delta H_{ab}/\Delta H_c = 0.2$ points toward an incommensurate order close to the commensurate one. This is illustrated in fig.4 where the apparition of a small transverse component for $h_{loc}$ is observed in the example of an incommensurate $(1/2-\varepsilon,0,l)$ order. We computed numerically the expected NMR spectra for $H_0//ab$ and $H_0//c$ in this model [18]. For small $R$, we find $R=5.0\varepsilon$. Comparison with experiment leads to $\varepsilon \approx 0.04$. The present study does not allow to rule out more complex wavevectors or small changes in the moment orientation, but the order of magnitude found for $\varepsilon$ would remain similar. Neutron scattering experiments detect some magnetic weight at $(1/2,0,l)$ but do not report any incommensurability [15]. However the small $\varepsilon$ deviation from the $(1/2,0,l)$ peak is beyond the q-space experimental resolution of the data reported for Co dopings above $4\%$. More generally, this anisotropy of the NMR linewidth in the frozen state appears a very sensitive probe of the incommensurability, and for a given model, is an effective tool to decide for the actual value of the wavevector coordinates.

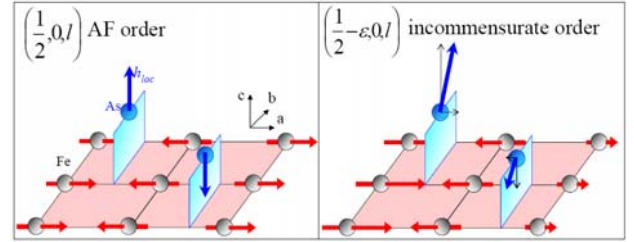

Fig. 4: (Color online) *Left panel* : the commensurate AF order in $BaFe_2As_2$ induces two values for the local field $h_{loc}$ on $^{75}As$ nucleus along $c$ axis only. *Right panel* : any incommensurate order such as the example displayed distributes $h_{loc}$ value along $c$ and along $(a,b)$ (moments are arbitrary scaled and only As atoms above the Fe layer have been shown).

The electronic properties of the Fe layer can also be probed through the spin lattice relaxation time $T_1$ which is linked to the low energy spin fluctuations through the dynamic spin susceptibility $\chi''(q,\omega)$:

$$1/T_1T \propto \sum_q |A_{hf}(q)|^2 \chi''(q,\omega_{NMR})/\omega_{NMR}$$

where $A_{hf}(q)$ is the $^{75}As$ hyperfine form factor. The $T_1$ were determined on the central line of the spectrum using standard saturation-recovery method and usual fitting procedures. The ratio $1/T_1T$ plotted versus temperature in fig.3 shows a first divergence at $T_{SDW}$ simultaneous to the appearance of the strong internal field. This is a signature of the critical slowing down of spin fluctuations before their freezing. A second divergence is observed just below $T_C$. The actual value of $T_C$ was checked *in-situ* under the applied field $H_0$ by monitoring the detuning of the NMR coil due to superconductivity (lower panel of fig.3). Similar $1/T_1T$ enhancement was also reported in the superconducting state of $CaFe_2As_2$ under pressure [14] and reveals a change in the dynamics of the system due to superconductivity. It is probably linked with the vortices present in the SC phase in this experiment. These two $1/T_1T$ peaks are further evidence that dynamic features due both to SDW and superconductivity are experienced on the same Fe sites. Finally, the decrease of $1/T_1T$ at low temperature is due to the superconducting gap development. It has been argued that this gap would display nodes in the coexisting regime only, which should be reflected by the T-dependence of $T_1$ [5]. This cannot be decided from present data because of the additional vortex effects on dynamics. Nuclear Quadrupole Resonance which does not require any magnetic field would be more appropriate but is very hard to achieve in Ba(Fe,Co)$_2$As$_2$ because of the much lower and more distributed values of the quadrupolar frequency compared to LaOFeAs.

The evolution versus carrier doping of the SDW state and the coexistence with SC can be understood in an itinerant picture. In the undoped semi-metal, commensurate nesting between identical hole and electron pockets leads to a commensurate SDW. Co doping increases the size of the electron pocket so that nesting deteriorates, i.e. $T_{SDW}$ and the moment get smaller, and incommensurability appears. The decrease of the moment shown in fig.1 follows well that of the transition temperature $T_{SDW}$, which is in favor of such an itinerant scenario, compared to an atomic localized magnetic point of view where the moment should not vary so much and reach so small values. At intermediate doping, Vorontsov et al. showed that this SDW may coexist with SC if $T_{SDW}/T_C$ is large enough and if the SDW is incommensurate [3]. Our results are compatible with this model, which could also explain why the hole-doped Ba$_{1-y}$K$_y$Fe$_2$As$_2$ shows no coexistence but segregation [9], the ratio $T_{SDW}/T_C$ being twice smaller for K-doping than for Co-doping. However, it is not clear why this deterioration of the magnetic order and the coexistence is not observed in La(O,F)FeAs, despite its large $T_{SDW}/T_C$ [6]. Furthermore, photoemission measurements of the Fermi Surface topology and band behavior reveal a more complex situation than a simple SDW gap opening due to hole-electron nesting [19]. It has been argued that such a simple itinerant nesting picture is indeed not sufficient and the magnetic order could be partially local because of large Hund's coupling [4]. This debate remains open, and precise determination of how the SDW evolves with doping among the various pnictide families is needed.

Regardless the origin of the magnetism, the coexistence itself demonstrated here is in essence different from that encountered in high T$_C$ cuprate or organic superconductors. In the latter componds, SC and AF seem indeed to be competing. When coexistence is observed using μSR or NMR, it is suspected to consist of one dimensional AF-stripes or spatial nano or micrometer segregation [11]. In our study, the homogeneous coexistence is underlying how these pnictides open a route towards original quantum states which remain to be fully understood. A study byJulien et al. (Europhys. Lett. 87 (2009) 37001) reaches similar conclusions about coexistence, but could not detect any NMR signal in the magnetic state.

We acknowledge for fruitfull discussions H. Alloul, F. Bert, F. Bouquet, V. Brouet, M. Gabay, D. Jérome, H. Luetkens, and P. Mendels.